# Magnetic ordering phase transition and abnormal brittleness in dilute Fe-Mn solid solution

Wei Liu[a,b], Yunfeng Liang[c,b,*], Xiangyan Li[b], Yichun Xu[b], Yange Zhang[b], Wenliang Li[d], Q. F. Fang[b], Caetano R. Miranda[e,*], Chuan-Lu Yang[a], C. S. Liu[b], Xuebang Wu[b,*]

[a]*School of Physics and Optoelectronics Engineering, Ludong University, Yantai 264025, China*

[b]*Key Laboratory of Materials Physics, Institute of Solid State Physics, HFIPS, Chinese Academy of Sciences, Hefei 230031, China*

[c]*Department of Systems Innovation, School of Engineering, The University of Tokyo, Tokyo 113-8656, Japan*

[d]*College of Energy Engineering, Xinjiang Institute of Engineering, Urumqi 830091, China*

[e]*Instituto de Física, Universidade de São Paulo, CP 66318, São Paulo, SP 05315-970, Brazil*

*Corresponding authors

Email address: Yunfeng Liang: liang@sys.t.u-tokyo.ac.jp, Caetano R. Miranda: crmiranda@usp.br, Xuebang Wu: xbwu@issp.ac.cn

**Abstract**

Experiments showed that solute Mn in bcc iron is in antiferromagnetic (AFM) coupling with iron neighbours below 2 at.% Mn, but is in ferromagnetic (FM) coupling at higher concentrations. Surprisingly, although Mn is an important alloying element in high-strength steels, it induces brittleness just at around 2 at.% Mn and higher concentrations. However, the mechanisms for the magnetic ordering phase transition and the abnormal brittleness remain unclear. Based on magnetism-constrained/unconstrained calculations and *ab initio* molecular dynamics simulations within density functional theory, we show that while the AFM phase prevails at low Mn contents, the FM phase becomes dominant at 1.85 at.% Mn and elevated temperatures. Our results suggest that the AFM-FM phase transition with increasing Mn concentration can be ascribed to the thermal effect. Furthermore, we find that the brittleness of the Fe-Mn alloys at intermediate Mn content might be related to the stress variations within the grains accompanying the local magnetic ordering changes.

## 1. Introduction

Knowledge of the magnetism-related properties of iron-based alloys is significant in condensed-matter physics and materials science [1–11]. Magnetism usually endows iron-based alloys with fascinating properties. One famous example is the $Fe_{0.65}Ni_{0.35}$ Invar alloy: whose lattice anharmonicity at finite temperatures is suppressed by the relaxation of magnetic structure [3,4]. In recent years, it has been found that the magnetism of some iron-based alloys is coupled with elastic and thermal properties, which is promising for future applications [5,6]. Element Mn is an important alloying addition to improve the mechanical and anticorrosion properties of steels [7,12–15]. Due to the half-filled 3*d* band, Mn is at the crossover between electron bonding behaviour and magnetic moment formation, and its 3*d* electron delocalization-localization transition can be induced by small perturbations [16]. Over decades, dilute body-centred cubic (bcc) Fe-Mn alloys have attracted considerable attention as prototypical model systems [9–14,17–35]. The available experimental studies show that solute Mn and the Fe neighbours are in an antiferromagnetic (AFM) coupling phase (AFM phase for short) at low concentrations but in a ferromagnetic (FM) coupling phase (FM phase for short) at concentrations higher than ~2.0 at.% Mn [18–22]. Unfortunately, it is always difficult to understand the origin of the AFM–FM transition. Predicting the transition theoretically is even more challenging, primarily due to the inadequate description of the complex magnetic interaction. Additionally, solute Mn leads to abnormal brittleness at concentrations approximately from 1.5 to 5 at.% [12,23–27], posing a challenge to understand the underlying mechanism.

Despite ongoing theoretical investigations, consensus has not been reached regarding the AFM-FM transition in dilute bcc Fe-Mn alloys. It was shown that the Fe-Mn phase diagram can be improved with better consistency with all available thermodynamic data by considering magnetic ordering energy [28,29]. Yet, part of the low-temperature phase equilibrium remains hypothetical. Early calculations using linear-muffin-tin-orbital (LMTO) Green's function method showed that the AFM and FM phases are both stable [30]. Density-functional theory (DFT) calculations within the local-density approximation (LDA) showed that the direction of the magnetic moment of Mn ($M_{\mathrm{Mn}}$) is sensitive to exchange-correlation potentials and parameter settings [30] and predicted the onset of the AFM–FM phase transition at much different Mn concentrations of 12 at.% and 2 at.%, respectively [32,33]. Recent DFT studies using the Perdew-Burke-Ernzerhof (PBE) functional within the generalized gradient approximation (GGA) pointed out that the AFM phase is the ground state. In contrast, the FM phase, referred to as a metastable state, is a shallow local minimum in the energy landscape versus $M_{\mathrm{Mn}}$ [13,14]. These DFT studies using the PBE functional presented different results on the AFM–FM phase transition [13,14,34]. From the $M_{\mathrm{Mn}}$ statistics on the disordered bcc Fe-Mn alloy, it was found that the configurations with lower average $M_{\mathrm{Mn}}$ are more energetically favourable even at 4.69 at.% Mn and inferred that a uniform AFM–FM transition at 2-3 at.% Mn might be an artefact from inadequate statistical sampling [13,33,34]. Another study using a special quasirandom structure model showed that the AFM–FM transition gradually progresses [14]: the proportion of FM-Mn solutes increases with increasing Mn concentration, making the average $M_{Mn}$ turn from negative to positive at ~6 at. % Mn.

Overall, it is still a formidable challenge for theoretical modelling to uncover the relative stability and the transition between AFM and FM phases in bcc Fe-Mn alloys. There is a longstanding debate in theoretical simulations on the phase transition in dilute Fe-Mn solid solution at approximately 2.0 at.% Mn [13,14,30–35]. To gain a clear picture of the phase transition, two tasks are crucial: (1) identifying the ground magnetic state of Fe-Mn solid solution at different Mn concentrations by a reliable simulation scheme and (2) investigating the thermal effect on the AFM–FM transition that has not been considered in the past theoretical studies to understand the experimental results at room temperature. Here, relying on an accurate description of the Fe-Mn interaction, the ground magnetic states at different temperatures and pressures were identified by magnetism-constrained/unconstrained DFT calculations, which provide new insights into the Fe-Mn phase diagram. Our study solves the longstanding debate in theoretical modelling on the AFM–FM magnetic ordering phase transition and agrees well with the experiments. The mechanism of thermally activated AFM–FM transition was suggested. This magnetic ordering phase transition involves electron transfer among 3*d* states of Mn. Analyzing stress variations induced by this phase transition, alongside phenomena such as Mn's clustering, segregation, and precipitation, it was suggested a possible correlation between this phase transition and the abnormal brittleness observed in diluted bcc Fe-Mn alloys.

## 2. Methods

DFT calculations in this work were spin-polarized as implemented in the Vienna *ab initio* simulation package (VASP) [36,37]. Most calculations were within the collinear magnetism approximation. Some additional noncollinear magnetism-constrained calculations were also performed. The interactions between ions and electrons were described by the projector-augmented wave (PAW) approach [38]. The energy cutoff for the plane-wave expansion of wave functions was set to 350 eV. We employed the Perdew and Wang (PW91) exchange-correlation functional within the GGA in all calculations [39] and the Vosko-Wilk-Nusair (VWN) method in interpolating the correlation part [40]. This scheme has been well tested for treating ferromagnetic iron alloys [10,41], and our test was stated in Supplementary Material (SM for short). A dense *k*-point mesh, which is necessary for calculating elastic moduli [42], was used in this study. The pseudopotential that treats semicore $3p^6$ as valence electrons was used for Mn, while the standard was used for Fe. Six different Mn concentrations from 0.4 to 2.08 at.% were considered by six systems with different supercells, where each has a single Mn atom in addition to Fe atoms. An additional $Fe_{106}Mn_2$ system in a tetragonal supercell was also employed. The dimensions of these supercells and the corresponding *k*-point meshes are listed in Table 1. The Monkhorst-Pack scheme was adopted for Brillouin-zone sampling in *k*-space. For the $3\times3\times4$ and $4\times4\times3$ bcc supercells, the length ratios of the [100] (or [010]) edge versus the [001] edge were set to 3⁄4 and 4⁄3, respectively. For the $3\times2\times2$ (48 atoms) and $3\times3\times3$ (108 atoms) tetragonal supercells [43], the length ratios of the [100] edge versus the [011] (or $[0\bar{1}1]$) edge were set to be $3/2\sqrt{2}$ and $\sqrt{2}/2$, respectively. The first-order

Methfessel-Paxton smearing method was employed in relaxing ions with constant volume, and the force convergence criterion was less than 0.003 eV/Å. Tetrahedron smearing with Blöchl corrections was employed in calculating electronic and magnetic properties. The constraint parameter increases in the $M_{Mn}$-constrained calculations until the penalty energy is less than one-hundredth of the energy convergence criterion of $10^{-5}$ eV. Two kinds of $M_{Mn}$-constrained methods were employed, termed as the 1st (collinear) and the 2nd (noncollinear), respectively. In the 1st method, the $M_{Mn}$ value was gradually increased from −2.8 to 2.2 $\mu_B$ while its orientation was collinear with that of the Fe neighbours. Conversely, in the 2nd method, the angle between the moments $M_{Mn}$ and $M_{Fe}$, confined within a certain plane, was adjusted stepwise while allowing for the relaxation of the magnitude of $M_{Mn}$. AIMD simulations were performed in a canonical ensemble with a Nosé thermostat for temperature control. The Verlet algorithm was used to integrate Newton's equations of motion with a time step of 4 fs. A gradual heating-up ramp was simulated from 50 to 500 K with a step of 50 K, and the duration was 2,000 fs at each temperature.

## 3. Results

**3.1. Mn concentration dependence of the energy landscape at equilibrium volume.** For fundamental knowledge about the ground state of the bcc Fe-Mn solid solution, we calculated the total energy of the Fe-Mn system at equilibrium volume as a function of the collinear $M_{Mn}$ under six different Mn contents based on the collinear magnetism-constrained DFT calculations. Figure 1a shows that the energy curve at 1.85 at.% Mn agrees well with the DFT curve calculated using PBE functional in Ref. 14. The relative energy $\Delta E$ in the $M_{\mathrm{Mn}}$ range from 0.3 to 0.8 $\mu_{\mathrm{B}}$ drops as the Mn content increases up to 1.85 at.%. Consequently, a shallow local minimum is observed on the right side of the lowest minimum, implying the electronic dichotomy of Mn. These results demonstrate that at ambient pressure, there is a possibility of a solute Mn with a positive $M_{\mathrm{Mn}}$ at a higher Mn concentration but little chance at a lower Mn concentration in the bcc Fe-Mn alloy. This is in perfect agreement with previous experiments, where the FM phase was only observed at ~2.0 at.% and higher Mn concentrations [18–22]. Based on magnetism-unconstrained DFT, we performed structural relaxations on the Fe-Mn system at each Mn content with an initial $M_{\mathrm{Mn}}$ of −2 and 2 $\mu_{\mathrm{B}}$, respectively. The findings are shown in Fig. 1b compared with experiments and previous simulation results [1,10,18–21]. Our calculations show solute Mn with positive $M_{\mathrm{Mn}}$ (FM phase) only at 1.39 at.% Mn and higher concentrations, which is consistent with experiments and the above prediction from the collinear magnetism-constrained DFT calculations.

Figure 1c shows a slight difference in lattice parameter ($a$) between the two phases but a more significant difference in bulk modulus ($B$). For example, the difference of $a$ is 0.0043 Å, and the difference of $B$ is 29.2 GPa at 2.08 at.% Mn. Herein, the considerable difference of $B$ is undoubtedly beyond the result of the slight volume difference but owing to the electronic structure change in the phase transition, which will be discussed in the latter part. The local lattice difference between these two phases can be conveniently displayed by comparing the distances of Fe neighbours to solute Mn: the first, third, fourth, and fifth nearest Fe neighbours are closer to the Mn ion in the FM phase than in the AFM phase, but the second nearest Fe neighbours are farther away from Mn, as shown in Table S1 in SM. These findings are consistent with the atomic relaxations due to Fe substitution with Mn in these two phases disclosed by previous DFT calculations using PBE functional [13], verifying the PW91 functional in describing the Fe-Mn system in this work (see SM 2 for more details).

**3.2. Temperature and pressure effects on the phase transition.** To explore the temperature effect on the stabilities of these two phases, AIMD simulations were performed on the $Fe_{53}Mn_1$ system at ~0 kBar (after Pulay stress correction, see SM 3). In our AIMD simulations, at each finite temperature, the appropriate volume was obtained from a tuning process by performing several trial simulations until the averaged pressure was within ±0.5 kBar. It shows that the lattice parameter increases linearly with increasing temperature up to 500 K. The linear thermal expansion coefficient was predicted to be $10.1\times10^{-6}$ $K^{-1}$, which is very close to the experimental value of $\alpha$-Fe, $11.7\times10^{-6}$ $K^{-1}$. Statistics on the relative distributions of absolute atomic displacements display a Maxwell-Boltzmann distribution at each temperature, as shown in Fig. S1 in SM. These curves fall into two sets: one consists of curves below 200 K and the other at 200 K and higher temperatures. There is a significant interval between the curves at 150 K and at 200 K, which might indicate a phase transition. This is probably because the system has much stronger anharmonic vibrations of the lattice at ≥200 K. At ≤150 K, both the total energy ($E$) and the external pressure ($P$) exhibit typical thermal fluctuations with very small amplitudes. However, at ≥200 K, the two quantities oscillate in resonance, exhibiting the same phase-angle and shape in their oscillograms. Meanwhile, the total magnetic moment ($M_{tot}$) oscillates in an opposite phase sequence. The case at 300 K is shown in Figs. 2(a-c). Remarkably, the amplitudes of the oscillating $E$, $P$, and $M_{tot}$ at ≥200 K are much larger than their respective values at ≤150 K. As in the case of $E$ shown in Fig. S2 in SM, the amplitude at ≥200 K is ~10 times greater than that at ≤150 K. In addition, the $M_{tot}$ values expand in the larger range as temperature increases and show a trend of two-humped distribution at ≥200 K (see Fig. 2d).

To explain these unusual behaviours at ≥200 K, we picked out configurations at the local extrema of $M_{tot}$ at 300 K and calculated $M_{Mn}$ in these configurations. It was found that the drastic change in $M_{Mn}$, approximately from −2.0 to 1.0 $\mu_{B}$, as shown in Fig. 2a, leads to a far more evident oscillation of $M_{tot}$ than a normal thermal fluctuation. The system takes AFM ($M_{Mn}<0$) and FM ($M_{Mn}>0$) phases in turn owing to thermal activation at ≥200 K, and the duration of either phase is ~150 fs. This means that the AFM and FM phase transition can be thermally activated at ≥200 K. Interestingly, valence (magnetic) fluctuations with durations in tens of femtoseconds have been found experimentally in $\delta$-Pu with dichotomous 5$f$ electrons [44]. We found a good correspondence between the $M_{Mn}$ and the cell volume ($V_{1nn}$) enclosed by the first nearest Fe neighbours of solute Mn as time evolves, as shown in Fig. S3 in SM. The oscillation can thus be understood as follows: when the $M_{Mn}$ is in the AFM state, e.g., at approximately 1,200±10 fs, the $V_{1nn}$ is large, which generates a positive pressure surrounding the local cell (as indicated by positive external pressure, $P$). This positive pressure enables $V_{1nn}$ to shrink. Consequently, at approximately 1,240–1,290 fs, the $M_{Mn}$ enters an FM state with a small $V_{1nn}$, which generates a negative pressure surrounding the local cell (as indicated by negative $P$). This negative pressure enables $V_{1nn}$ to expand, and then the $M_{Mn}$ enters an AFM state with a large $V_{1nn}$. These dynamics present an AFM–FM transition mechanism by contracting the local cell around the solute Mn atom and the reversed transition by expanding the local cell. Based on our understanding of the oscillation behaviour, we hypothesize that the oscillation may be perturbed by stochastic dynamics. To corroborate this hypothesis, we carried out AIMD simulations with a Langevin thermostat (see SM 4) using the last-step structure in the FM state, and it was found that the system stayed in the FM state most of the time (Fig. S4). Additionally, when we used the AFM state structure as the starting point, the system transformed into an FM state at approximately 20 fs (Fig. S5). Both simulations show the dominating duration of the FM state (Fig. S6). This means that the FM state is more stable upon perturbation. The result can be understood based on the total energy difference between the two phases (see below).

Since the $Fe_{53}Mn_1$ system has only one Mn atom, it takes only a magnetic coupling state between Fe and Mn atoms at a time in our AIMD simulations, which enables us to estimate the total energy difference between the two phases. From Fig. 2c, the total energy at an $M_{Mn}$ of approximately −2.0 $\mu_B$ in the AFM phase can be ~1.0 eV higher than that at an $M_{Mn}$ of approximately 0.8 $\mu_B$ in the FM phase, which is very different from the case at 0 K: the former is ~0.05 eV lower than the latter, as shown in Fig. 2e. This result indicates that the FM phase is the ground state of the bcc Fe-Mn alloy with Mn content ≥1.85 at.% at 200 K and higher temperatures. Statistics show that the sum of the square of atomic displacement in the AFM phase is more significant than that in the FM phase at 300 K, as shown in Fig. S7 in SM. This means that the anharmonic effect is more significant in the AFM phase than in the FM phase at the same temperature. We suggest that the higher total energy of the AFM phase than that of the FM phase at ≥200 K might be attributed to the larger deviation from the equilibrium configuration in the AFM phase than in the FM phase. By using the total energy difference between the two phases (~1.0 eV) and the Boltzmann distribution function, it is predicted that only one per $6.7 \times 10^{16}$Mn atoms takes negative $M_{\mathrm{Mn}}$ (AFM phase), overwhelmed in FM-Mn atoms in a real system at 300 K. This explains the experimental observations that only the FM phase exists at ~2.0 at.% Mn and higher concentrations [18–22].

To explore pressure effects on the stabilities of these two phases, the compression run was performed by decreasing the volume of the $Fe_{53}Mn_1$ system step by step at 0 K. As shown in Fig. 3a, the enthalpy in the FM phase becomes lower than that in the AFM phase as pressure increases at ~40.0 kBar (corresponding to a volumetric strain of −2.4%), indicating that the ground state shifts from the AFM to FM phase. A similar compression run was also performed on a $Fe_{106}Mn_2$ system, where two Mn atoms were randomly distributed in the bcc lattice. The resulting fitted enthalpy curve closely resembles that of the $Fe_{53}Mn_1$ system, as shown in Fig. 3a (fitted parameters listed in SM 5). This result is consistent with the finding that a compressive strain of −2% stabilizes the FM phase for the $Fe_{127}Mn_1$ system from previous DFT calculations [13]. However, it should be noted that the required pressure was as high as ~40.0 kBar. We remark that the temperature effect discussed above is more crucial to explaining the experimental data [18–22].

The total energies at different $M_{Mn}$ values were calculated using the collinear magnetism-constrained DFT for the $Fe_{53}Mn_1$ system with two compressive strains, $0.978V_0$ and $0.954V_0$. As shown in Fig. 3b, the local minimum of the total energy in the AFM phase ($M_{Mn} \sim -1.3\ \mu_B$) is almost equal to that in the FM phase ($M_{Mn}$ ~$0.7\ \mu_B$) at $0.978V_0$. At $0.954V_0$, there is no local minimum of the AFM phase but only one global minimum of the FM phase in the energetic curve versus $M_{Mn}$. These energy profiles under compressive strains (0.978 $V_0$ and 0.954 $V_0$) are in sharp contrast to that at $V_0$, supporting the FM phase as the ground state. The noncollinear magnetism-constrained DFT calculations were also performed at different volumes, and the energy variations with the angle $\theta$ between the moments $M_{Mn}$ and $M_{Fe}$ are displayed in Fig. 3c. Regardless of whether at the equilibrium volume or under hydrostatic compressive strain, the local minima of total energy are consistently located at $\theta$=0° and 180°, representing the FM and AFM phases, respectively. Notably, the total energy in the FM phase gradually drops with increasing compressive strain (downward along the blue vertical line in Fig. 3c), ultimately reaching its global minimum at 0.954 $V_0$. Therefore, it is reasonable to infer that the FM phase is the ground state beyond a certain pressure threshold. Furthermore, our findings suggest that collinear magnetism approximation can be applicable in the DFT modelling on the Fe-Mn solid solution system.

**3.3. Electronic structure changes in the phase transition.** To better understand the transition between the two phases, we compare the electronic structures of the $Fe_{53}Mn_1$ system in different phases at equilibrium volume at 0 K based on collinear magnetism-unconstrained DFT calculations. First, we examine the 3*d* density of states (DOS) of Mn. Figures 4a and b display the $t_{2g}$ and $e_g$ DOS curves in both phases. Herein, the DOS of the $t_{2g}$ band is the average of $3d_{zx}$, $3d_{yz}$, and $3d_{xy}$, while the DOS of the $e_g$ level is the average of $3d_{z^2}$ and $3d_{x^2-y^2}$. As one can see, the peaks in the downward branch of the $t_{2g}$ DOS shrink while the upward $e_g$ peak shifts in the low-energy direction to be below $E_F$ energy from the AFM to FM phase. The proportion of $e_g$ electrons in the total 3*d* electrons is 0.336 in the AFM phase but increases to 0.384 in the FM phase. Rahman *et al*. observed a similar trend of $e_g$ peak shift in a systematic study of the DOS curves of 3*d* solutes from Mn to Co in bcc iron based on all-electron calculations [9]. Second, we examine the differential electron density $\rho^{diff}$ in the (110) plane of solute Mn from the AFM to FM phase. As shown in Fig. 4c, there is a noticeable electron transfer from regions along the body diagonals ([$\bar{1}$11] and equivalent directions) to regions along the [001] direction around solute Mn. This electron transfer perfectly corresponds to the shrinking of the downward $t_{2g}$ band and the shift of the upward $e_g$ peak from above $E_F$ energy to the below (as sketched in Fig. 4d). Third, we examine the change in the electron localization function (ELF) [45] from the AFM to the FM phase. In the AFM phase, the ELF value of bonding attractors between Mn and the first nearest neighbours (1NNs) is ~10% less than that between Mn and the second nearest neighbours (2NNs). However, they are equal in the FM phase. Figure 4e shows the shrinkage of the ELF bonding domains between Mn and 2NNs in the FM phase compared to the AFM phase. This result reflects the relative increase in the ELF value of the bonding attractors between Mn and 1NNs, indicating strengthened bonding between Mn and 1NNs. Interestingly, the Fe-Mn bond strengthening can be explained within classical valence bond theory, assuming that the electrons in the antibonding $t_{2g}$ bands are transferred into $e_g$ bands [1]. The $3d^5$ electrons of Mn behave much differently in the two phases, exhibiting the electronic dichotomy feature. The substantial $B$ increase from the AFM to FM phase is reasonable because it is mainly attributed to the strengthened bonding between Mn and 1NNs. Meanwhile, the increased attraction between Mn and 1NNs compresses the $Fe_{53}Mn_1$ system, leading to a slight contraction of the lattice parameter from the AFM to FM phase. The weakened bonding between Mn and 1NNs in the AFM phase might be responsible for the stronger anharmonic effect at temperatures ≥200 K. In addition, we found no significant difference in the charge, electronic DOS, and magnetism of 1NNs and 2NNs between the two phases, seemingly indicating a weak interplay between solute Mn and the neighbouring Fe ions during the phase transition. This result is consistent with the conventional idea of an isolated localized magnetic state of solute Mn in bcc iron [2,18,20]. That is, the presence of Mn in bcc iron does not cause a strong frustration in the magnetic moment orientation of Fe neighbours.

## 4. Discussion

The calculated energy landscapes with the collinear $M_{Mn}$ at different Mn contents predict that only AFM Fe-Mn coupling phase exists at ≤1.39 at.%. AIMD simulations show that the AFM phase is the ground state, and the FM phase is metastable at temperatures below 200 K. Conversely, at high temperatures in Fe-Mn alloy with 1.85 at.% Mn, the FM phase is the ground state. Compression simulations at 0 K further indicate the FM phase is the ground state in bcc Fe-Mn alloy with 1.85 at.% Mn at pressures higher than ~40 kBar. These results contribute to the broader understanding of the Fe-Mn phase diagram. There are common features of the electronic structure change among this AFM–FM transition and some other phase transitions. The $e_g$ shift is similar to the $a_{1g}$ shift in the metal-insulator transition of corundum $V_2O_3$ and the $3d_{xz}$+$3d_{yz}$ shift in the isostructural phase transition of the hexagonal $YCo_5$ [46,47]. Our stochastic dynamics simulations at 300 K on the $Fe_{53}Mn_1$ system presented a thermally activated two-phase fluctuation with a dominant occurrence of the FM phase. It is noteworthy that, to the best of our knowledge, valence/magnetism fluctuation phenomena have only been found in 5*f* metals Cf and Pu by experimental or theoretical studies [44,48–50].

As shown in Fig. S8 in SM, our calculated Pugh's shear-to-bulk modulus ($G/B$) ratio [23,51,52] predicts that the FM phase possesses superior ductility to the AFM phase. The decreased $G/B$ ratio combined with the increased $B$ (Fig. 1c) might partly account for the improved hardness and ductility of the steels at ≥5.0 at.% Mn [12]. The previous theoretical efforts on the embrittlement of iron alloys, including studies on the effect of electronic configuration of solutes on cohesive energy and solute effects on grain boundaries, may inadequately elucidate the abnormal brittleness observed in bcc Fe-Mn alloy at concentrations approximately from 1.5 to 5.0 at.% Mn [23,53–56]. Herein, the abnormal brittleness is characterized by a considerably increased brittle–ductile transition temperature, decreased fracture toughness relative to pure $\alpha$-iron, and transcrystalline fractography [23–27]. In recent theoretical studies, the brittleness of dilute Fe-Si alloy at 10.9 at.% Si was attributed to the instability of a large narrow peak at Fermi energy, and the magnetic ordering effect was suggested to be significant in the temperature dependence of brittleness [57]. We hypothesized that the abnormal brittleness of bcc Fe-Mn alloy may also correlate with this AFM–FM transition. Our calculations on the bcc $Fe_{126}Mn_2$ (1.56 at.% Mn) system show that the $M_{Mn}$ is significantly influenced by the distance between solutes Mn, and correspondingly, the $B$ and $a$ of the system are altered (see Fig. S9 in SM). This implies that Mn clustering changes the elastic and structural properties of different regions in a grain. The segregation/precipitation of Mn leads to its lower concentration in the regions near grain boundaries, which might induce the transition from FM to AFM phase in some Mn-diluted regions near grain boundaries. From Fig. S9a, the lattice parameter ($a$) change due to the AFM-to-FM transition or Mn clustering can be ~0.003 Å, corresponding to a volumetric strain of ~0.32%. This volumetric strain can induce stress of ~610 MPa, far greater than the tensile strength of bcc Fe-Mn ferrite steels: ~300 MPa [15]. Consequently, the changes in equilibrium volume accompanying such magnetic ordering phase transitions engender stress concentration regions in some parts of the grains. This analysis suggests that fracture may occur within these stress concentration regions, showing transcrystalline fractography, and consequently, resulting in a significant reduction in fracture toughness observed in Charpy impact tests at room temperature [24–26]. On the other hand, despite the thermally activated nature of the AFM–FM transition in the regions containing approximately 2.0 at.% Mn being localized and instantaneous, its occurrence cannot be omitted in practical systems, even over timescales extending to seconds. This transition induces time-varying stress with amplitude over 800 MPa (8.0 kBar, see Fig. 2b) in the grains, thereby exacerbating the brittleness. This study is expected to inspire further endeavours aimed at employing *ab initio* molecular dynamics to study the temperature effect on the magnetic interactions of transition-metal-based materials.

**Data and materials availability**

All data needed to evaluate the conclusions in the paper are presented in the paper and/or the Supplementary Materials.

**Acknowledgements**

We appreciate the valuable discussions with Prof. Zhao-Rong Yang at the High Magnetic Field Laboratory, Chinese Academy of Sciences, and Prof. Xue-Feng Sun at the University of Science and Technology of China (USTC). This work is supported by the National Key Research and Development Program of China (Grant Nos.:

2022YFE03110002, 2017YFA0402800, and 2016YFE0120900) and the National Natural Science Foundation of China (Nos.: 52325103, 12192282, 51871207, 52071314, 12374232, and U1832206). C.R. Miranda acknowledges the financial support provided by the Brazilian Ministry of Science and Technology for collaborative research between China and Brazil and the Brazilian funding agencies National Council of Scientific and Technologic Development (CNPq) and Fundação de Amparo à Pesquisa do Estado de São Paulo (FAPESP) (Grant 2017/02317-2).

**Author contributions**

C.S.L., Y. L, X.W., C.R.M., and W.L. designed the research; W.L. carried out simulations; W.L., Y.L., X.W., and C.R.M. analyzed the data, discussed the results, and wrote the manuscript; all authors commented on the manuscript.

**Competing interests**

The authors declare that they have no competing interests.

**Additional information**

**Supplementary information**

The online version contains supplementary material available online associated with this paper.

**Correspondence and requests for materials** should be addressed to Y. Liang, C. R. Miranda, and X. Wu.

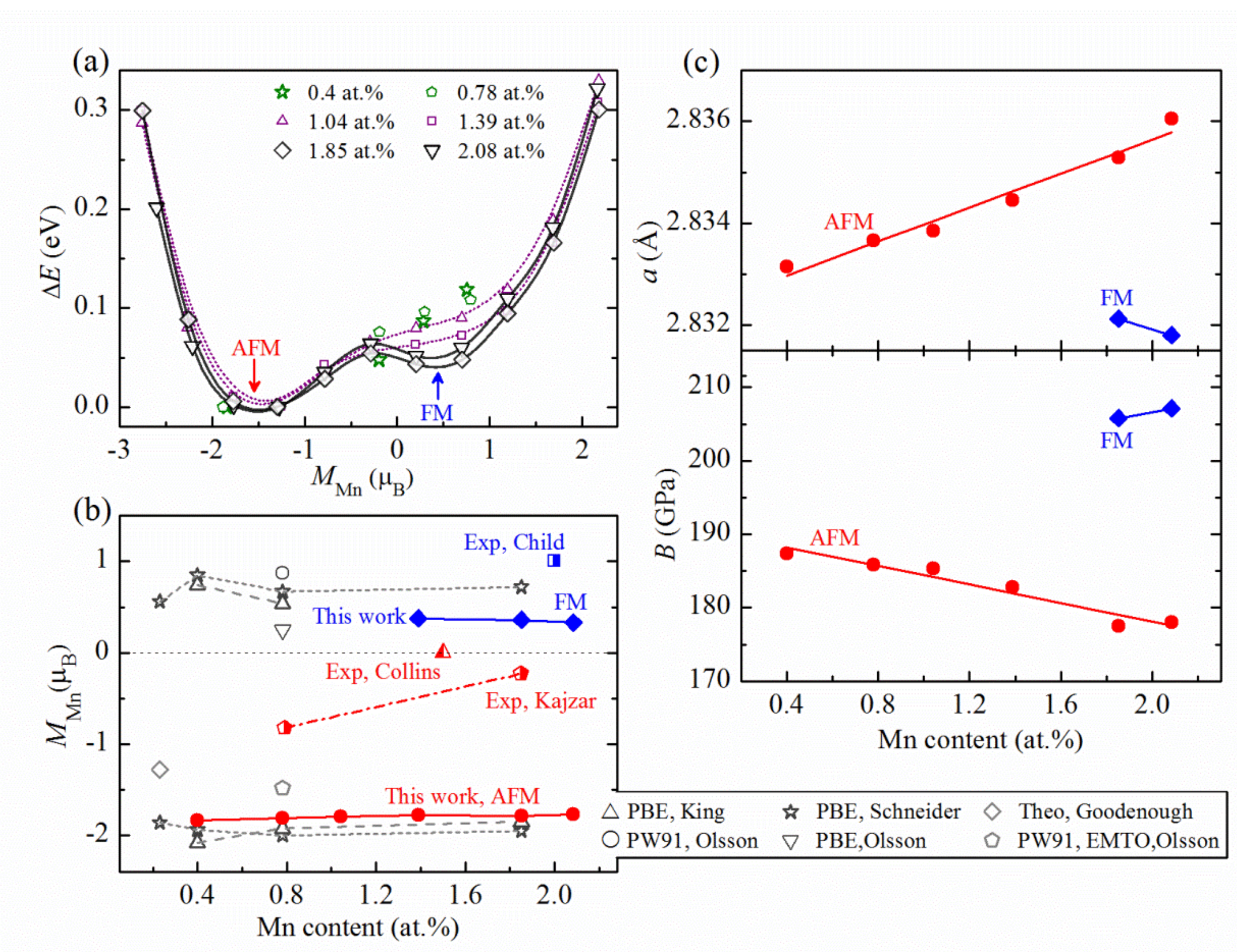

**Fig. 1. Energetic, magnetic, structural, and elastic features of AFM and FM phases at 0 K.** (a) Relative total energy ($\Delta E$) of the Fe-Mn system as a function of the collinear magnetic moment of Mn ($M_{\mathrm{Mn}}$) at Mn contents from 0.4 to 2.08 at.%. The local minima of the AFM and FM phases are indicated by red and blue arrows, respectively. (b) $M_{\mathrm{Mn}}$ versus Mn content in the Fe-Mn system. The cited data are from Refs. (1, 10, 13,14,18–22). The theoretically predicted $M_{\mathrm{Mn}}$ in the dilute limitation in Ref. (1) is arbitrarily placed at 0.23 at.% Mn. (c) Variations in the lattice parameter (upper) and bulk modulus (lower) with Mn content. Red lines are linear fits to the AFM data. Note that the FM state at 1.39 at.% Mn is only a critical state due to the relatively flat energy profile, as shown in (a); hence, it is not included in (c).

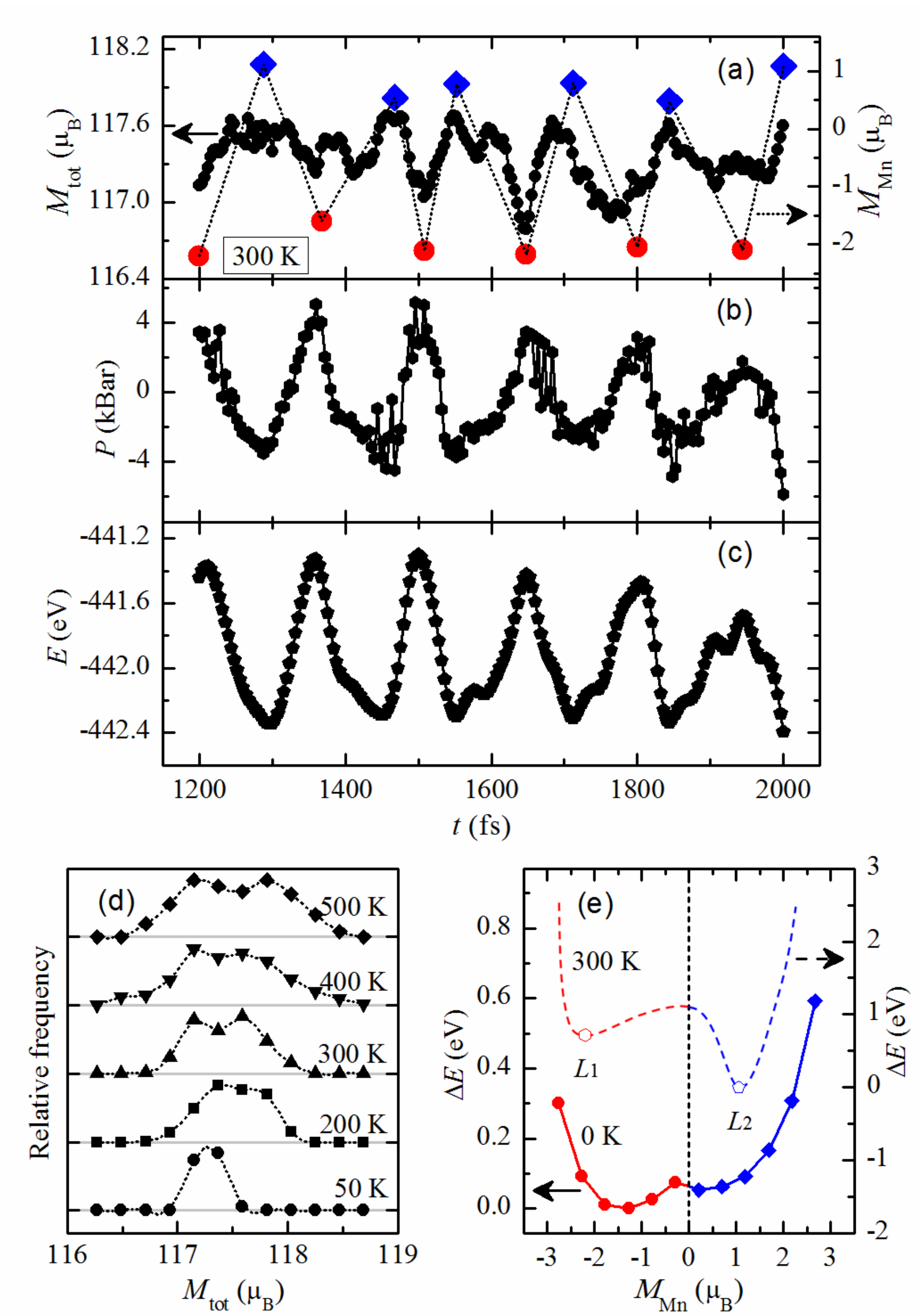

**Fig. 2. Two-phase oscillations in the $Fe_{53}Mn_1$ system at 300 K and magnetic and energetic features of the two phases.** Evolution of (a) total magnetic moment ($M_{tot}$) of the $Fe_{53}Mn_1$ system and magnetic moment ($M_{Mn}$) of solute Mn, (b) external pressure, and (c) total energy versus time in AIMD simulations at 300 K. (d) Relative occurrence frequency of different $M_{tot}$ values at temperatures from 50 to 500 K. (e) Relative total energy ($\Delta E$) as a function of $M_{Mn}$ at 0 K and 300 K. Herein, the curve at 300 K is a schematic. The two local minima marked $L_1$ and $L_2$ are from AIMD.

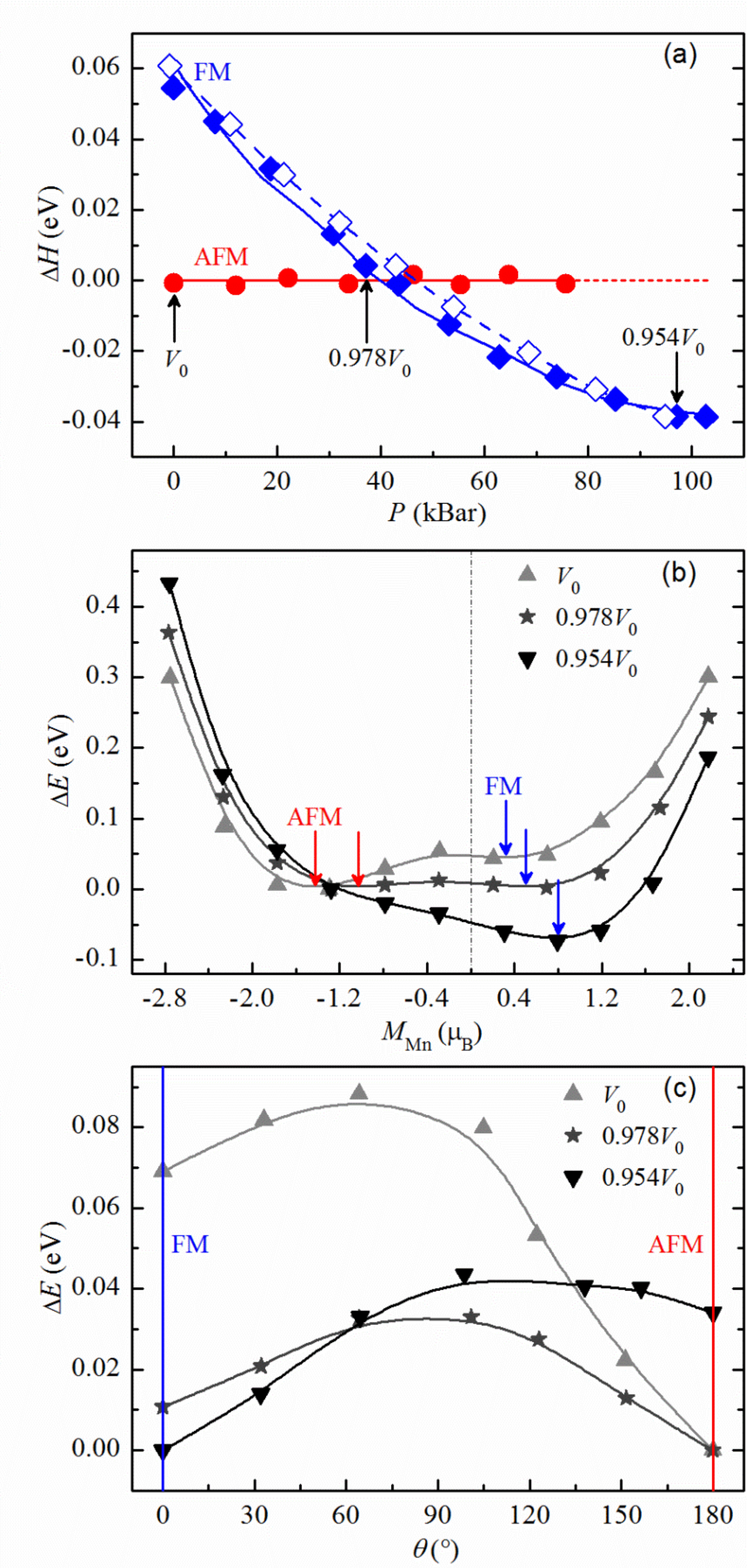

**Fig. 3. Ground-state switch from AFM to FM phase under compression.** (a) Enthalpy difference ($\Delta H$) between the FM phase and AFM phase of the Fe-Mn system with 1.85 at.% Mn as a function of pressure. For the convenience of the display, the fitting and extrapolated enthalpy data in the AFM phase, represented by red solid and red short-dashed lines, respectively, are set to zero at each pressure. Red circles represent the *ab initio* enthalpies in the AFM phase. Blue solid diamonds represent the *ab initio* enthalpies of the $Fe_{53}Mn_1$ system in FM phase, and the blue solid line represents the corresponding fitting results. Blue hollow diamonds represent the *ab initio* enthalpies (divided by 2) of the $Fe_{106}Mn_2$ system in FM phase, and the blue dash line represents the corresponding fitting results. The black arrows mark different volumes, and $V_0$ is the equilibrium volume in the AFM phase. (b) Relative total energy ($\Delta E$) as a function of the magnetic moment of Mn ($M_{\mathrm{Mn}}$) at three different volumes. The red and blue arrows denote the energetic minima of AFM and FM phases, respectively. (c) $\Delta E$ as a function of the angle between $M_{\mathrm{Mn}}$ and that of Fe atoms. Red and blue vertical lines denote the angles in AFM and FM phases, respectively.

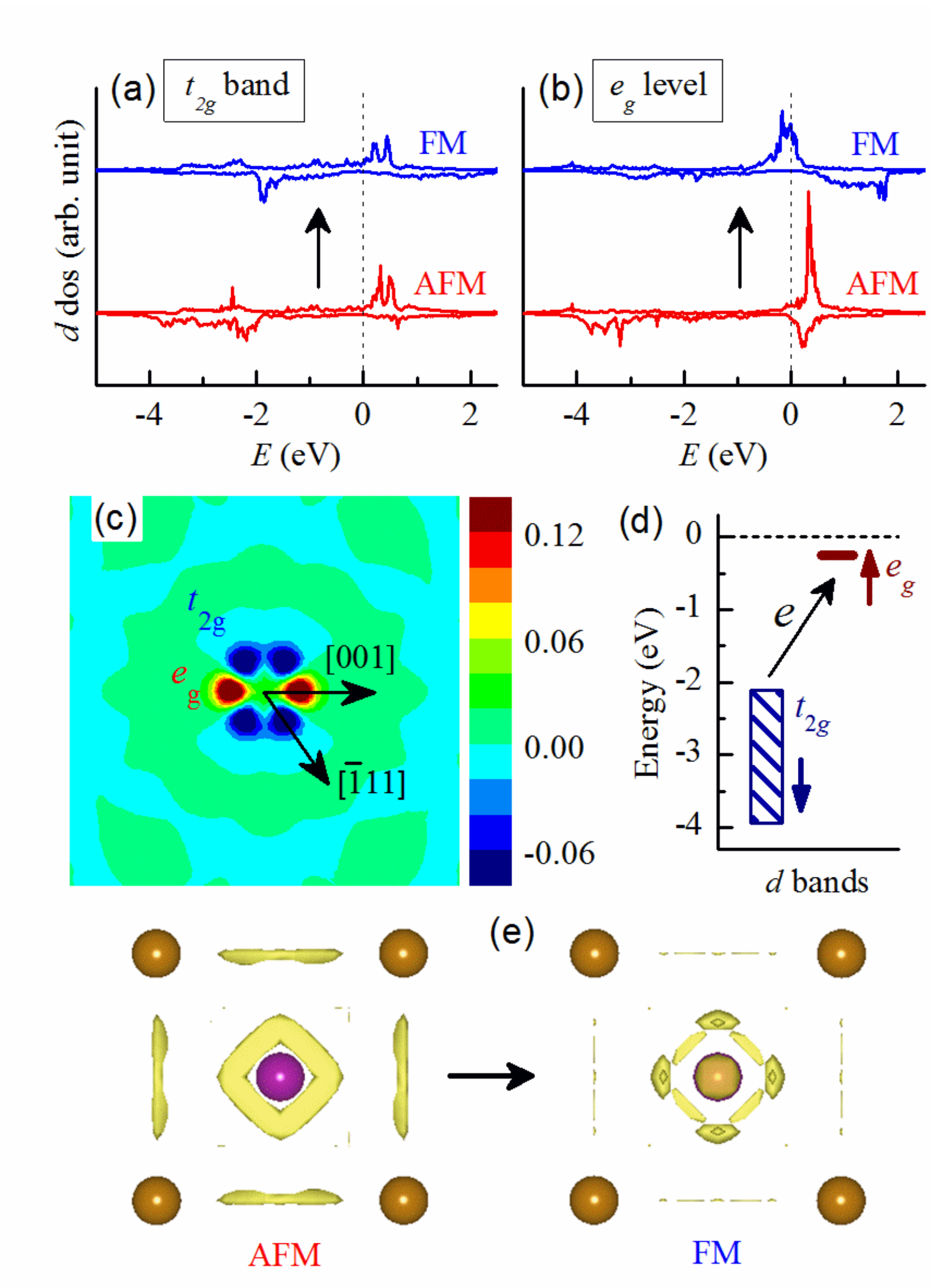

**Fig. 4. Electronic structure change from AFM to FM phase transition at 0 K.** Spin-polarized DOS curves of (a) 3*d* $t_{2g}$ and (b) 3*d* $e_g$ states of Mn from the AFM to FM phase in the $Fe_{53}Mn_1$ system. (c) Differential electron density of Mn on the (110) plane from the AFM to FM phase. The unit of electron density is $e/Å^3$. Solute Mn is placed at the centre. (d) Schematic of electron transfer within 3*d* states. (e) ELF domains around solute Mn (purple ball at the centre) from the AFM to FM phase. Deep yellow balls represent the first nearest Fe neighbours, and the second nearest Fe neighbours are not shown. The ELF value is set to be the attractors between Mn and 1NNs.

**Table 1.** Solute Mn contents (at.%) of seven supercells, dimensions, and corresponding *k*-point meshes. All *k*-point meshes are Γ-centered. For the first five supercells, dimensions are in multiples *l*, *m,* and *n* of a two-atom bcc cell along the [100], [010], and [001] directions, respectively; for the last two supercells, dimensions are in multiples *l*, *m,* and *n* of a four-atom tetragonal cell [see FIG. 1(a) in Ref. (43)] along the [100], [011] and [0$\bar{1}$1] directions, respectively. Herein, *l*, *m,* and *n* are integers.

| at.% | supercell | dimension | *k*-point mesh |
|---|---|---|---|
| 0.40 | $Fe_{249}Mn_1$ | 5×5×5 | 4*4*4 |
| 0.78 | $Fe_{127}Mn_1$ | 4×4×4 | 5*5*5 |
| 1.04 | $Fe_{95}Mn_1$ | 4×4×3 | 5*5*7 |
| 1.39 | $Fe_{71}Mn_1$ | 3×3×4 | 7*7*5 |
| 1.85 | $Fe_{53}Mn_1$ | 3×3×3 | 7*7*7 |
| 1.85 | $Fe_{106}Mn_2$ | 3×3×3 | 7*5*5 |
| 2.08 | $Fe_{47}Mn_1$ | 3×2×2 | 7*10*10 |